\documentclass[fleqn,twoside,twocolumn,nofootinbib,showkeys]{revtex4} 
\usepackage[sec,nocpr]{ujp} 
\begin{document}
\title[The dynamics of general Bianchi IX model near the cosmological singularity]
{THE DYNAMICS OF GENERAL BIANCHI IX MODEL\\ NEAR THE COSMOLOGICAL SINGULARITY}%
\author{S.L.~Parnovsky}
\affiliation{Astronomical Observatory of Taras Shevchenko Kyiv National University}
\address{Observatorna str., 3, 04058, Kyiv, Ukraine}
\email{parnovsky@knu.ua}

\udk{530.12} \pacs{04.20.-q} \razd{\seci}

\autorcol{S.L.~Parnovsky}

\setcounter{page}{1}%

\begin{abstract}
Half a century ago, Belinsky and Khalatnikov proposed a generic solution of the Einstein equations near their cosmological 
singularity, based on a generalization of the homogeneous model of Bianchi type IX. Consideration of the evolution of the 
most general non-diagonal case of this model is greatly simplified if it is assumed that, when approaching the singularity 
$t=0$, it reduces to the so-called asymptotic dynamics, at which inequality (6) holds. It has been suggested that this 
inequality continues to be true from the moment of its first fulfilment up to the singularity of space-time. We 
analyze this assumption and show that it is incorrect in the general case. However, it is shown that there is always a time 
$t_0$, after which this assumption becomes true. The value of $t_0$ is the smaller the less is the degree of non-diagonality 
of the model. Some details of the behaviour of the non-diagonal homogeneous model of Bianchi type IX are considered at the 
stage of asymptotic dynamics of approaching the singularity.
\end{abstract}

\keywords{general relativity, cosmology, singularity, general solution.}

\maketitle

\section{Introduction}\label{s1}

The Belinskii-Khalatnikov-Lifshitz (BKL) conjecture is thought to be a generic solution to the Einstein
equations near spacelike cosmological singularity \cite{BKLspace}. By generic solution one means, roughly
speaking, the solution that is stable against perturbations of the initial conditions defining the dynamics
and the set of these conditions is of nonzero measure in the space of all possible initial conditions.
The general solution should include the maximal possible number of ``physically arbitrary'' functions of all three
space coordinates, i.e. four functiones for vacuum space and eight ones for space filled with matter fields \cite{LK,LL}.

The considered problem is closely related to the general problem of the existence of spacetime singularities.
On one hand, the existence of such solutions may mean that there are some intrinsic problems in general relativity
as it is believed that a physical theory should be free of singularities.  On the other hand,
the singularity theorems argue that singularities are integral parts of
general relativity (see \cite{Senovilla:2014gza} and references therein). It is commonly
expected that quantization of general relativity may ``heal'' the classical singularities.

Quantization of the BKL scenario should be preceded by quantization of the Bianchi IX model.
It seems to be a natural strategy as the BKL scenario was obtained by analysing the
dynamics of the Bianchi IX spacetime \cite{BIX}. For analyses of the dynamics
of the Bianchi IX universe, carried  out within the dynamical systems method, we recommend  \cite{BogNov} (diagonal
case of the space 3-metric) and \cite{Bog} (nondiagonal case). Quantization
of gravitational systems is known to be quite complicated due to the nonlinearity of the
dynamics and an existence of dynamical constraints (relations among degrees of freedom).
One usually tries to identify special cases minimizing these difficulties. In the context of
cosmological singularities, one tries to find the dynamics that addresses just the singularity
problems without additional complications.

As the result, some hope has been associated with the simplified dynamics considered
long time ago. The first results were published in 1971 by V.A. Belinskii, I.M. Khalatnikov, and M.P. Ryan in the preprint \cite{BKRyan}. 
However, it did not appear in the form of an article, but became a part of the paper \cite{Ryan} and the 
basis for some further research and contains a number of important conclusions about the asymptotical dynamics of the system
\cite{Belinski:2014kba, Czuchry:2012ad}.

We see that this work is related to a problem that arose half a century ago and has not yet been solved due to its complexity. 
During this time, the concepts of dark energy (DE) and the associated accelerated expansion of the Universe have been firmly 
established in the general theory of relativity. Convincing evidence for the existence of DE is the dependence of the redshift 
and photometric distance for type Ia supernova explosions. However, the existence of DE either in the form of a cosmological 
constant or in the form of a dynamic DE can affect the generic solution near singularities. This can be seen in the example of 
singularities, which are the source of not only gravitational, but also scalar fields \cite{PSc1,PSc2}. In this article, we do 
not consider the effect of DE, limiting ourselves to the influence of matter and radiation. We can say that we consider the 
problem in the very form in which it was posed 50 years ago. Note that it remains relevant in some models of dynamic DE, in 
which its influence disappears near the singularity.

The aim of this paper is the study of some details of the asymptotical dynamics including the
correctness of assumption, on which the paper \cite{Ryan} is based.
In the next section we recall, following \cite{BKRyan,Ryan,Belinski:2014kba}, the way the dynamics of the 
general (non-diagonal) Bianchi IX universe transforms into the asymptotic dynamics.  
It is based on making the special assumption concerning some of the terms defining the dynamics. 
Afterwards, we study the asymptotical dynamics and investigate whether this assumption
is correct in the sense that it can be satisfied in a continuous way in the evolution
of the system from the moment of its imposition until approaching the singularity.
This determines whether the simplified system of equations can represent the general dynamics
in the asymptotic regime near the singularity.

The BKL solution was extended later
to the  timelike singularity \cite{BKLtime}, and merged to general scenario including both
singularities \cite{BKLboth}. There are some differences between both cases, but the general
dynamics near timelike and spacelike singularities is the same \cite{PP}. In this paper we consider
only a generic solution near spacelike cosmological singularity, but its result could be easy
transferred to the case of a timelike one.

\section{Exact dynamics and its asymptotic form}\label{s2}

In what follows we use the terminology and notation of Refs.
\cite{BIX,Ryan}.  The general form of a line element of the nondiagonal Bianchi IX
model in the synchronous reference system is
\begin{equation}\label{dd1}
ds^2 = dt^2 - \gamma_{ab}(t)\mathbf{e}^a_\alpha \mathbf{e}^b_\beta dx^\alpha
dx^\beta ,
\end{equation}
where Latin indices $a,b,\ldots$ run from $1$ to $3$ and label
frame vectors, and Greek indices $\alpha,\beta,\ldots$ take values
$1,2,3$ and concern space coordinates, and where $\gamma_{ab}$ is
a spatial metric.

The homogeneity of the Bianchi IX model means that the three
independent differential 1-forms $e^a_\alpha dx^\alpha$ are
invariant under the transformations of the isometry group of the
Bianchi IX model, i.e. $e^a_\alpha (x) dx^\alpha = e^a_\alpha
(x^\prime) {dx^\prime}^\alpha$, so $e^a_\alpha$ have the same
functional form in old and new coordinate systems. Each
$e^a_\alpha dx^\alpha$ cannot be presented in the form of a total
differential of a function of coordinates. 
Three frame vectors $\mathbf{e}^{a}$ of the Bianchi type IX
homogeneous space are presented in \citep{LL,Sch} in the form
\begin{equation}
\begin{array}{l}
\label{e5a}
\mathbf{e}^{(1)}=\mathbf{l}=(\sin z,-\cos z\sin x,0),\\
\mathbf{e}^{(2)}=\mathbf{m}=(\cos z,\sin z\sin x,0),\\
\mathbf{e}^{(3)}=\mathbf{n}=(0,\cos x,1).
\end{array}
\end{equation}

The spatial metric
$\gamma_{ab}(t)$ is a $3\times3$ matrix with all nonvanishing
elements to be determined. We put
\begin{equation}
\label{g}
\boldmath {\gamma}=\mathbf{O}^{-1}\mathbf{\Gamma} \mathbf{O},
\end{equation}
where $\mathbf{O}(t)$ is the orthogonal matrix parametrized in the standard way by
the Euler angles $\psi, \theta, \varphi$ and $\mathbf{\Gamma}(t)=diag(\Gamma_1(t),\Gamma_2(t),\Gamma_3(t))$.
So, the matrix $\gamma_{ab}$ is comletely described by six time dependent parameters 
$\psi, \theta, \varphi,\Gamma_1,\Gamma_2,\Gamma_3$.

We examine an approximate solution of Einstein equations
for the Bianchi IX model with the metric (\ref{dd1})
near the singularity which corresponds to $t=0$.
We can restrict ourselves to the vacuum case because the energy-momentum tensor components for a
hydrodynamically moving matter give negligible contribution to the dynamics as $t\to 0$.
The influence of DE in the general case can be significant, but in this article we assume that it can be neglected.

It was shown  \cite{BIX} that near the cosmological
singularity the {\it general} form of the metric $\gamma_{ab}$
should be considered. Consequently, one cannot 
diagonalize the metric for all values of time. After
making use of the Bianchi identities, freedom in the rotation of
the metric $\gamma_{ab}$ and frame vectors $e^a_\alpha$, one
arrives at the system of equations
specifying the dynamics of the nondiagonal Bianchi IX model.

They have simpler form if we redefine the cosmological time variable $t$ as follows: $ dt = \sqrt{\gamma}\; d\tau$,
where $\gamma$ denotes the determinant of $\gamma_{ab}$. However,  the equations including 
$\dot{\psi}, \dot{\theta}, \dot{\varphi},\ddot{\Gamma}_1,\ddot{\Gamma}_2,\ddot{\Gamma}_3$, where ``dot'' denotes $d/d\tau $, 
are still quite
complicated. The explicit form of these equations is presented in \cite{Ryan} [see Egs. (2.14)--(2.20)] or in
\cite{Belinski:2014kba} [see Eqs. (34)--(38)].

These equations include the constant $C$ which is important for our purposes, so we give its definition.
One can get, from the Einstein equations \cite{Ryan}, the condition for the tensor 
$\kappa_a^b=\gamma^{bc}\dot{\gamma}_{ac}$ in the form
\begin{equation}\label{c1}
\kappa_a^b\varepsilon^a_{\phantom{a}bc}=C_c \, .
\end{equation}
Here $\varepsilon_{abc}$ is the Levi-Civita symbol and $C_c$ are three
constants which can be considered as components of a three-dimensional vector. Choosing the third
coordinate axis along the direction of this vector we get
\begin{equation}\label{c2}
C_1=C_2=0,\quad C_3=C.
\end{equation}
The diagonal case corresponds to $C=0$ and to an absence
of rotation, i.e. $\dot{\varphi}=\dot{\theta}=\dot{\psi}=0$.

Belinsky, Khalatnikov, and Ryan \cite{BKRyan} proposed a conjecture, which 
greatly simplifies the general form of the dynamics near the singularity, if 
one accepts that whenever an inequality $\Gamma_1 > \Gamma_2 > \Gamma_3$ holds 
true, it becomes a strong one:
\begin{equation}\label{strong}
  \Gamma_1 \gg \Gamma_2 \gg \Gamma_3
\end{equation}
(the order of indices in \eqref{strong} is unimportant and depends on initial 
conditions). The primary goal of this article is to verify if such a conjecture 
is valid.

If the condition \eqref{strong} is satisfied, $\theta$, $\varphi$, and $\psi$ 
tend to some constant values $\theta_0$, $\varphi_0$, and $\psi_0$ near the 
singularity. The set of equations is substantially reduced to
(see, \cite{BKRyan,Ryan} for more details)
\begin{equation}\label{b1} 
\begin{array}{l}
\frac{d^2 \ln a }{d
\tau^2} = \frac{b}{a}- a^2,~~~~\frac{d^2 \ln b }{d \tau^2} = a^2 -
\frac{b}{a} + \frac{c}{b},\\
\frac{d^2 \ln c }{d \tau^2} = a^2 -
\frac{c}{b},
\end{array}
\end{equation}
where $a,b,c$ are functions of time $\tau$, satisfying the
constraint:
\begin{equation}\label{b2}
\begin{array}{l}
\frac{d \ln a}{d \tau}\;\frac{d \ln b}{d \tau} + \frac{d \ln a}{d
\tau}\;\frac{d \ln c}{d \tau} + \frac{d \ln b}{d \tau}\;\frac{d
\ln c}{d \tau}\\
= a^2 + \frac{b}{a} + \frac{c}{b},
\end{array}
\end{equation}
and where
\begin{equation}\label{b3}
\begin{array}{l}
a :=\Gamma_1,~~b:= \Gamma_2 C^2 \cos^2 \theta_0,\\
c := \Gamma_3
C^4 \sin^2\theta_0 \cos^2\theta_0 \sin^2\psi_0 \, .
\end{array}
\end{equation}
We assume $\sin \theta_0\neq0$, $\cos \theta_0\neq0$ and $\sin\psi_0\neq0$. 
Note that the definition of $b$ and $c$ includes the parameter $C$ and the
diagonal case corresponds to $C=0$. Thus, there is no simple reduction of this 
nondiagonal case to the diagonal one.

Belinsky, Khalatnikov, and Ryan \cite{BKRyan,Ryan} argue that the condition \eqref{strong} is
satisfied during the evolution of the system towards the cosmological singularity.
Therefore, in the asymptotic regime near the singularity, the spatial 3-metric
can be determined by the solution to the equations \eqref{b1}--\eqref{b2} without the loss of the generality.

\section{The asymptotic dynamics near the singularity}

Introducing new variables $\xi=\ln(a^2)$, $\eta=\ln(b/a)$
and $\zeta=\ln(c/b)$ we get Eqs. \eqref{b1} in the form
\begin{equation}\label{e1}
\ddot\xi=2(e^{\eta}-e^{\xi}), \, \ddot\eta=2(e^{\xi}-e^{\eta})+e^{\zeta}, \, \ddot\zeta=e^{\eta}-2e^{\zeta}.
\end{equation}
Naturally, the set of equations \eqref{e1} is much simpler than the general one, but it
requires the condition \eqref{strong} to be valid. Is this condition carried out all the time? From \eqref{strong} and \eqref{b3} we 
have the relations
\begin{equation}\label{e0}
\begin{array}{l}
\Gamma_1/\Gamma_2=C^2 \cos^2 \theta_0(a/b)=C^2 \cos^2 \theta_0e^{-\eta}\gg 1,\\ 
\Gamma_2/\Gamma_3=C^2 \sin^2 \theta_0 \sin^2\psi_0e^{-\zeta}\gg 1.
\end{array}
\end{equation}
The only way to satisfy the inequalities \eqref{e0} over the entire range of $\eta$ and $\zeta$ is
to require
\begin{equation}\label{crit}
\begin{array}{l}
C\gg \cos^{-1} \theta_0e^{\eta_{\max}/2},\\ 
C\gg \sin^{-1} \theta_0 \sin^{-1}\psi_0e^{\zeta_{\max}/2},
\end{array}
\end{equation}
where $\eta_{\max}$ and $\zeta_{\max}$ are the maximal values of $\eta$ and $\zeta$.
The inequality \eqref{crit} sets the lower limit for the value of $C$.

We can also consider \eqref{crit} as a restriction on the maximal values of the functions $\eta$ and $\zeta$. Let us choose 
certain values $\eta_0$ and $\zeta_0$ depending on $C$, $\theta_0$, and 
$\psi_0$ in such a way that the inequality \eqref{e0} is fulfilled at 
$\eta=\eta_0$, $\zeta=\zeta_0$ but violated at larger values of $\eta$ or $\zeta$.
This reduces the inequality \eqref{strong} to the conditions
\begin{equation}\label{u1}
\eta<\eta_0,~~~~\zeta<\zeta_0,
\end{equation}
which have to be satisfied during evolution until the singularity.
So we are interested in the dependence of $\eta$ and $\zeta$ on $\tau$.

\begin{figure}
\includegraphics[width=\columnwidth]{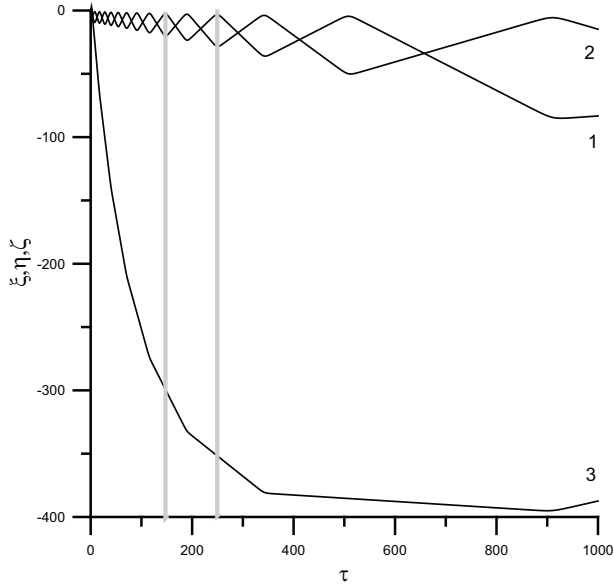}
\caption{An example of dynamics of $\xi=\ln(a^2)$ (line 1), $\eta=\ln(b/a)$ (line 2) and $\zeta=\ln(c/b)$
(line 3) as functions of $\tau$. Two vertical grey lines corresponds to two successive maxima of $\xi$}
\label{f3}
\end{figure}
\begin{figure}
\includegraphics[width=\columnwidth]{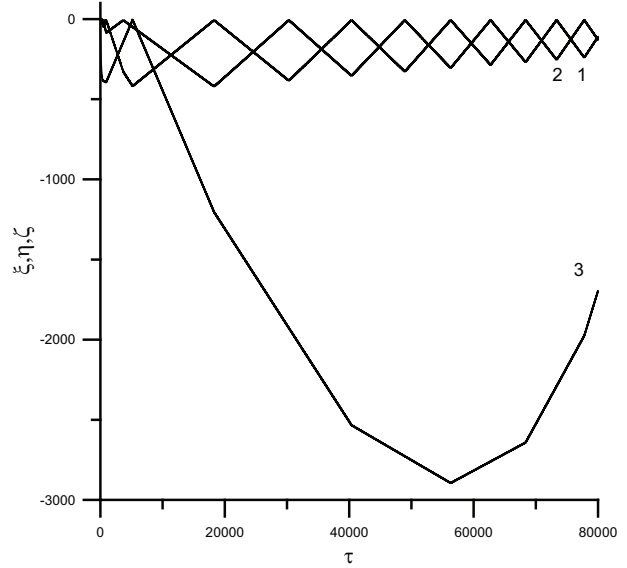}
\caption{The same as Fig. \ref{f3} but on larger time interval}
\label{f2}
\end{figure}
In Fig. \ref{f3} we plot the results of numerical solution of \eqref{e1} with rather arbitrarily chosen initial values and derivatives 
satisfying the condition \eqref{b2}. This is only an illustration of the dynamics of functions $\xi$, $\eta$ and $\zeta$. 
Next we'll study it analytically. In the meantime, we note that maxima of all functions are achieved at values close to zero.
However, the local minimum of $\zeta$ is achieved at $\zeta\approx -400$. But 
this is not the global one. If we extend the plot to larger $\tau$ region we obtain Fig. \ref{f2} with values of $\zeta$ below -1900.
The value of $\exp(\zeta)$ varies by more than 850 orders in this plot. Should have our calculations continued, we would 
find a local minimum lower that the first one and the function $\zeta$ would start to increase.

If we use the condition \eqref{u1} alone and choose $\zeta_0=-250$, then it starts to fulfil at $\tau\approx 100$ and continue to
be valid until $\tau\approx 3000$, after which the condition \eqref{strong} is 
violated. This means that the simplified system of equations \eqref{b1} -- \eqref{b2} cannot represent the general
dynamics at $\tau >100$.

Can these conditions be fulfilled until the singularity? We need some analitycal study to answer this question.

The functions $a(\tau),b(\tau)$, $c(\tau)$ and $\xi(\tau)$, $\eta(\tau)$, $\zeta(\tau)$
undergo complex oscillations which cannot be described analytically with
all details. An evolution towards the singularity takes a finite interval of cosmological time $t$, but an infinite interval
of the evolution parameter $\tau$. An infinite number of oscillations occur during this time interval. They are separated
by the so-called Kasner epochs when the space-time is similar to the well-known Kasner metric \citep{Kas}
\begin{equation}
\label{eqq14}
\mathrm{d} s^{2} = \mathrm{d} t^{2} - t^{2p_{1}} \mathrm{d} x^{2} - t^{2p_{2}} \mathrm{d} y^{2} - t^{2p_{3}}\mathrm{d} z^{2},
\end{equation}
where Kasner indices $p_{i}$ satisfy the conditions
\begin{equation}
\label{eqq16}
p_{1} + p_{2} + p_{3} = 1,\quad {p_{1}}^{2}
+{p_{2}}^{2} +{p_{3}}^{2} = 1.
\end{equation}
Thus one of them is negative and other two are positive. More precisely, they are subject to the inequalities  
\begin{equation}\label{e16}
-\frac{1}{3}\leq p_{\mathrm{min}}\leq 0\leq p_{\mathrm{mid}}\leq \frac{2}{3}\leq p_{\mathrm{max}}\leq 1, 
\end{equation}
where the Kasner indices $p_{i}$ are ranked in value. The index $p_{\mathrm{max}}$ is the largest one,  
$p_{\mathrm{min}}$ is minimal one, which is always negative, and $p_{\mathrm{mid}}$ lies between them.
Kasner epochs correspond to almost linear sections in Figs. \ref{f3} and \ref{f2}.

Consider a region of space bounded by the borders with constant values of space coordinates $x^{\alpha}$.
Its volume $dV=\gamma^{1/2}dx^1dx^2dx^3 \propto abc$ depends on $\tau$. Let us introduce the function 
\begin{equation}\label{e10}
\begin{array}{l}
Q(\tau)=\frac{d\ln V}{d\tau}=(\ln(a))^{\cdot}+(\ln(b))^{\cdot}+(\ln(c))^{\cdot}\\
=\frac{3}{2}\dot{\xi}+2\dot{\eta}+\dot{\zeta}.
\end{array}
\end{equation}
It satisfies the equation
\begin{equation}\label{e10a}
\dot{Q}=e^{\xi}.
\end{equation}
\begin{figure}
\includegraphics[width=\columnwidth]{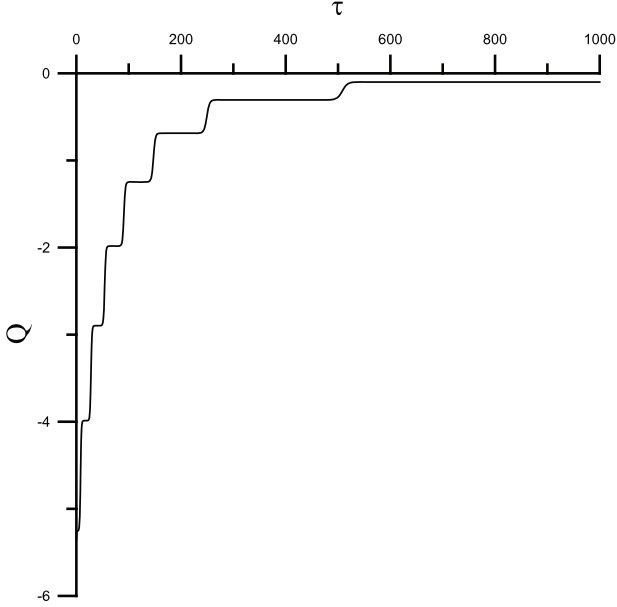}
\caption{The function $Q(\tau)$ for the functions $\xi,\eta,\zeta$ from Fig. \ref{f3}} 
\label{f5}
\end{figure}

The function $Q(\tau)$ for the functions $\xi$, $\eta$ and $\zeta$ from Fig. \ref{f3} is plotted in Fig. \ref{f5}. One can see that it 
looks like stairs with flat steps. We denote its value at steps by $\Lambda_i$, where $i$ is the number of step, starting with 0.
We do not {\it a priori} know the sign of $\Lambda_0$, 
which depends on initial values of $\dot\xi,\dot\eta,\dot\zeta$ at $\tau=0$. We can choose the direction towards the
singularity from two possibilities, namely the directions along or opposite to the $\tau$ increase. The natural choice 
is the direction in which $V$ decreases, so we get $V\to 0$ at $t\to 0$.
If $Q(0)>0$ or in different words $\Lambda_0>0$, $\tau \to -\infty$. If $\Lambda_0<0$, $\tau \to \infty$.

Let's apply the approach successfully used in the papers \cite{BKLspace,LL}. On many intervals of $\tau$ variation, the right-hand 
sides of the equations \eqref{e1} and \eqref{e10a} can be neglected in comparison with the main terms in the left-hand sides. These 
are the so-called Kasner epochs. During each Kasner's epoch $\ln (a)$, $\ln (b)$ and $\ln (c)$
are linear functions of $\tau$: 
\begin{equation}\label{e2}
\begin{array}{l}
Q=\Lambda_i,\,\xi=K_i\tau+\mathrm{const}, \,\eta=L_i\tau+\mathrm{const},\\ 
\zeta=M_i\tau+\mathrm{const},\,\ln (a)=A_i\tau+\mathrm{const},\\ 
\ln (b)=B_i\tau+\mathrm{const},\,
\ln (c)=D_i\tau+\mathrm{const}.
\end{array}
\end{equation}
Here $i$ be a number of maxima of 
functions $\xi(\tau),\eta(\tau),\zeta(\tau)$ between the current and the initial values of $\tau$. This number is used as an 
index numbering consecutive Kasner epochs. Each of them is characterized by its own set of Kasner indices.
The rule for changing the indices for adjacent Kasner epochs and
other details are described in \cite{Ryan,Belinski:2014kba}. 

It is clear that
\begin{equation}\label{e12}
\begin{array}{l}
K_i=2A_i, \,L_i=B_i-A_i, \,M_i=D_i-B_i,\\
\Lambda_i=A_i+B_i+D_i=1.5K_i+2L_i+M_i.
\end{array}
\end{equation}
For the Kasner epoch, the dependence of $a,b,c$ on time is similar to metric \eqref{eqq14}, so for it $\tau = \ln(t) + const$. 
From \eqref{eqq14} and \eqref{e10} we can express
\begin{equation}\label{e14}
A_i=\Lambda_i\cdot (p_1)_i,\,B_i=\Lambda_i\cdot (p_2)_i,\,D_i=\Lambda_i\cdot (p_3)_i,
\end{equation}
where $(p_1,p_2,p_3)_i$ is the $i$-th set of Kasner indices $p_1,p_2,p_3$ satisfying the conditions \eqref{eqq16}.

Kasner epochs are separated by transition epochs, in which one of the functions $\exp(\xi(\tau))$, $\exp(\eta(\tau))$, or 
$\exp(\zeta(\tau))$ entering the right-hand sides of the equations 
\eqref{e1} and/or \eqref{e10a} increases so much that it can no longer be neglected. 
Indeed, among the terms on the right side of \eqref{e1} in a general case there is one that increases faster than the terms on the left 
side when approaching the singularity.
However, one can neglect the rest of the terms 
in the right-hand sides and solve the resulting system of equations.
In all cases, the function that plays the role of perturbation in the right-hand sides first increases, reaches its maximum value, 
then begins to fall, and the solution passes into the next Kasner epoch. The change of epochs is clearly visible in Figs. \ref{f3} and 
\ref{f2}. Transition epochs correspond to the areas of $\tau$ variation around the maxima of one of three functions $\xi,\eta,\zeta$.

Having a basic understanding of the character of the asymptotic dynamics near the singularity we can consider the behavior of these 
functions during transition epochs. We consider three different cases separately.
The approximate solutions near each maximum have the following forms, where $S,K,L,M=$const.

Near a maximum of $\xi(\tau)$ at $\tau=T_{\xi}$ we get
\begin{equation}\label{e3}
\begin{array}{l}
\xi=2\ln(S)-2U(\tau),\, \eta=2U(\tau)+L\tau+\mathrm{const},\\
\zeta=M\tau+\mathrm{const},\,U(\tau)=\ln(\cosh(S(\tau-T_{\xi})),\\ 
Q=\mathrm{const}+S\tanh(S(\tau-T_{\xi})),
\end{array}
\end{equation}
Near a maximum of $\eta(\tau)$ at $\tau=T_{\eta}$ we get
\begin{equation}\label{e4}
\begin{array}{l}
\eta=2\ln(S)-2U(\tau),\, \xi=2U(\tau)+K\tau+\\
\mathrm{const},\quad \zeta=U(\tau)+M\tau+\mathrm{const},\\
U(\tau)=\ln(\cosh(S(\tau-T_{\eta}))), \, Q=\mathrm{const},
\end{array}
\end{equation}
Near a maximum of $\zeta(\tau)$ at $\tau=T_{\zeta}$ we get
\begin{equation}\label{e5}
\begin{array}{l}
\zeta=2\ln(S)-2U(\tau),\quad \xi=K\tau+\mathrm{const},\\
\eta=U(\tau)+L\tau+\mathrm{const},\quad Q=\mathrm{const},\\
U(\tau)=\ln(\cosh(S(\tau-T_{\zeta}))).
\end{array}
\end{equation}
We started from the assumption that we can neglect all terms in the right-hand side of \eqref{e1} except one containing a function of the 
variable which reaches a maximum. It is easy to verify that this condition is satisfied for each of the solutions 
\eqref{e3} -- \eqref{e5} during the transition epoch. After its completion $U(\tau)$ reaches a linear asymptotic and the system switches 
to the next Kasner epoch.  
Matching the solutions \eqref{e3} -- \eqref{e5} with the solutions \eqref{e2} at both sides of the maximum we obtain the rules of
changing of $K_i,L_i,M_i,\Lambda_i$ on passing through transition epochs in the forms
\begin{equation}\label{e6}
\begin{array}{l}
K_{i+1}=-K_i,\,L_{i+1}=L_i+2K_i,\, M_{i+1}=M_i,\\
\Lambda_{i+1}=\Lambda_i+K_i=\Lambda_i(1+2(p_1)_i),
\end{array}
\end{equation}
\begin{equation}\label{e7}
K_{i+1}=K_i+2L_i,\,L_{i+1}=-L_i,\, M_{i+1}=M_i+L_i,
\end{equation}
\begin{equation}\label{e8}
K_{i+1}=K_i,\,L_{i+1}=L_i+M_i,\, M_{i+1}=-M_i
\end{equation}
for maxima of $\xi$, $\eta$ and $\zeta$, respectively.
So, the function $Q(\tau)$ is constant except the regions of maxima of $\xi$, 
where it looks like a step function with the charecteristic $\tanh$-like shape. One can see an example of it in Fig. \ref{f5}.
The value of $|Q|$ monotonically decreases to zero when approaching the singularity. This decrease is irregular. Intervals with 
almost constant $Q=\Lambda_i$ are connected with the sections with a sharp change of $Q$. These changes are caused by tending $\tau$
to maxima of $\xi$ and are described by the hyperbolic tangent $\tanh(S(\tau-T_{\xi}))$. Its 
absolute value decreases at maxima of $\xi$, where $(p_1)_i<0$, but the sign 
remains the same. The values of $\Lambda$ do not change at maxima of $\eta$ or 
$\zeta$ according to \eqref{e10a}.
After the $n$-th such maximum we obtain
\begin{equation}\label{e19}
\Lambda_n=\Lambda_0\prod_{j=1}^n(1+2(p_1)_j)\xrightarrow [n \to \infty]{}0.
\end{equation}

We can obtain the values $\xi_{\max}$, $\eta_{\max}$ and $\zeta_{\max}$ for the $i$-th maximum
\begin{equation}\label{e190}
\begin{array}{l}
\xi_{\max}=2\ln(L_i/2)=2\ln(\Lambda_i(p_1)_i),\\
\eta_{\max}=2\ln(L_i/2)=2\ln(\Lambda_i(p_2-p_1)_i/2), \\
\zeta_{\max}=2\ln(M_i/2)=2\ln(\Lambda_i(p_3-p_2)_i/2),\\
\exp(\eta_{\max}/2)_i\propto \Lambda_i,\,\exp(\zeta_{\max}/2)_i\propto \Lambda_i.
\end{array}
\end{equation}
and substitute them into \eqref{crit}. Taking into account the decreasing of $|\Lambda|$ at each maximum of $\xi$ we
see than the condition \eqref{crit} is fulfilled near the singularity for any nonzero value of $C$ at $\sin \theta_0\neq0$, 
$\cos \theta_0\neq0$ and $\sin\psi_0\neq0$.

In terms of inequalities (\ref{u1}) this can be formulated in the following way. The local maxima of $\eta$ and $\zeta$
become smaller as $\tau$ increases, and sooner or later these inequalities 
begin to be satisfied for all subsequent values of $\tau$ (we assume that $\Lambda_0<0$ and the singularity corresponds to $\tau=\infty$ as in Figs. 
\ref{f3}-\ref{f5}). Let us denote the value of $\tau$ where this happens as 
$\tau_0(C,\theta_0,\psi_0)$. The greater the value of $C$ for the same 
$\theta_0,\psi_0$, the smaller the value of $\tau_0$.
So, the inequality \eqref{strong} is satisfied during the evolution of the set of equations (\ref{b1},\ref{b2}) of 
simplified asymptotical dynamics near the singularity and therefore this 
assumption is not contradictory in principle. However, as we just demonstrated, 
this conjecture is only valid if applied within the $\tau$ interval between 
$\tau_0$ and the singularity.

\section{Some relations for the asymptotical dynamics near the singularity}

Let us estimate some properties and parameters of the asymptotical dynamics near the singularity.
First of all we are interesting in a relationship between the cosmological time $t$, the new time coordinate $\tau$ and the 
volume $V\propto abc$. Unfortunately, a steplike form of the function $Q(\tau)$ complicates this problem. 

Consider one step or plateau of this function. It is located between two adjacent peaks of $\xi$, which are separated by 
the interval $\Delta\tau_i$. We mark two such peaks in Fig. \ref{f3} by grey lines. During this interval the function $\xi$ 
decreases from $\xi_{max}$ to $\xi_{min}$ and then increases almost to the initial value at the rate \eqref{e2}.
We know the value of $\xi_{max}$ from \eqref{e190}. If we consider the typical situation 
$\zeta\ll \xi,\eta$, then \eqref{e1} gives us
\begin{equation}\label{e100}
\ddot\xi+\ddot\eta=0,\quad\xi+\eta=R\tau+Y,\quad R,Y=const.
\end{equation}
We can find the value of $R$ from \eqref{e2} and \eqref{e14}. 
A local minimum of $\xi$ almost coincides with a local maximum of $\eta$ given by \eqref{e190}. This gives us the
value of $\xi_{min}$. As a result, we get an estimation 
\begin{equation}\label{e101}
\begin{array}{l}
\Delta\tau_i\approx\frac{2\ln(\Lambda_i(p_2+p_1)_i/2)-\Lambda_i(p_2+p_1)_i\tau-Y}{\Lambda_i(p_1)_i}\\
\xrightarrow [\tau \to \infty]{}s_i\tau,\quad 
s_i=-1-\left(\frac{p_2}{p_1}\right)_i.
\end{array}
\end{equation}
Here $s_i = u_i^{-1}$, where $u>0$ is a parameter used to write Kasner indices in 
parametric form \cite{LL}
\begin{equation}\label{e102}
\begin{array}{l}
p_1=\frac{-u}{1+u+u^2},\;p_2=\frac{1+u}{1+u+u^2},\; p_3=\frac{u(1+u)}{1+u+u^2},
\end{array}
\end{equation}
which changes its value when one Kasner era replaces another \cite{LL,Ryan}.

In \eqref{e101} we deal with terms depending on $\tau$ and on $\Lambda_i$. The latter increases according to \eqref{e19}.
Consider the case of very big $\tau_0$, corresponding to small $t$.
In a very rough approximation, $\ln(\Lambda_i)$ depends linearly on $i$, i.e. the number of maxima of $\xi$. 
Assuming that the term proportional to $\tau$ is the main one, we get the estimation \eqref{e101} for very large 
values of $\tau$. It yields that $\Delta\tau_i\propto \tau$ and increases with $\tau$ growth. One can see this in 
Figs. \ref{f3} -- \ref{f5}. So, our assumption proved correct and $\ln(\Lambda_i)$ grows slower than $\tau$.

After the interval $\Delta\tau_i$ from \eqref{e101} we have the next step and the value of $Q$ changes by 
\begin{equation}\label{e103}
\Delta Q =\Lambda_{i+1}-\Lambda_i=2(p_1)_i\Lambda_i
\end{equation}
according to \eqref{e6}. We remind that $(p_1)_i<0$.

One could approximate the function $Q$ by the sum of Heaviside step functions, but this leads to cumbersome expressions.
However, it is possible to get the same qualitative result by using a very 
rough approximation by replacing each of the steps by the line with a slope 
$\Delta Q/\Delta\tau$ and $\lambda_i$ by $Q$, which yields a much simpler 
differential equation
\begin{equation}\label{e104}
\dot{Q}=W_i\frac{Q}{\tau},\quad W_i=\frac{2(p_1)_i}{s_i}=\left(\frac{-2u^2}{1+u+u^2}\right)_i.
\end{equation}
So $-2<W_i<0$. We can very roughly suppose that $W_i=W=const<0$ and get ($P,D$=const)
\begin{equation}\label{e105}
Q\approx P\tau^W,\,V\propto \exp (D\tau^{1+W}),\, D=\frac{P}{1+W}.
\end{equation}
One can regard $W$ as a kind of a mean value of $W_i$, so $-2<W<0$. We see that at $-1<W<0$ the volume $V$ vanishes at 
singularity and at $-2<W<-1$ it tends to some finite value. Only the first case has a physical meaning and we shall consider 
it. The interval of cosmological time $t$ between a point with coordinate 
$\tau=\tau_1$ and the singularity at $\tau=\infty$ is equal to
\begin{equation}\label{e106}
\Delta t\propto\int_{\tau_1}^{\infty}\exp (D\tau^{1+W}/2) d\tau
\end{equation}
with $D<0$. This integral reduces to the incomplete gamma function and is finite at $-1<W<0$, so the singularity corresponds to some 
finite cosmological time and we can choose it as the origin of the coordinate $t$. Near the singularity i.e. for large $\tau_1$ we have
the asymptotic behaviour $V\propto t^2$.

Now let us consider how the value of $\tau_0$ after which the condition \eqref{strong} is satisfied depends on value of $C$. According to
\eqref{crit} and \eqref{e190} we can rewrite this condition in the form
\begin{equation}\label{e107}
C>C_0=FQ(\tau_0).
\end{equation}
Here $F$ is a factor depending on $\theta_0,\psi_0$ and also on what we mean by the term ``much larger'' in \eqref{strong}. So,
near the singularity  
$Q(\tau_0)\propto\tau_0^W\propto C$ and $\tau_0\propto C^{1/W}$. The value $t_0$ corresponding to $\tau_0$ is estimated as
\begin{equation}\label{e108}
t_0\propto\exp(ZC^{\omega})
\end{equation} 
with $\omega=1+W^{-1}<0$ and $Z=const<0$. This dependence is very strong for small values of $C$, i.e. for almost diagonal metrics.

\section{Conclusion}

The main result of our analyses is finding that there exists an instant of time, $\tau_0$, in the {\it asymptotic} evolution 
of the {\it general} dynamics  of the Bianchi IX universe towards the singularity at $\tau \rightarrow +\infty$, such that for $\tau > \tau_0$, 
the strong inequality \eqref{strong} is satisfied\footnote{One can choose the initial conditions for the dynamics in such a 
way that the singularity occurs at $\tau \rightarrow - \infty$, in which case the condition \eqref{strong} is fulfilled, but for $\tau < \tau_0$.}. Therefore, the asymptotic dynamics is {\it self-consistent}, but for $\tau > \tau_0$.

The value $\tau_0$ depends on the constant $C$, which characterizes the 
non-diagonality of the metric. The smaller its value, the greater the value of 
$\tau_0$. Taking the limit $C \rightarrow 0$, which leads to the diagonal case 
when applied to the exact dynamics of the Bianchi IX model, does not make sense 
at the level of the asymptotic dynamics.

We are aware that the considerations concerning the asymptotic dynamics might be devoid of any physical meaning because of possible 
quantum effects. All our analyses make the assumptions that quantum effect can be neglected. The importance of quantum effects 
will be examined elsewhere. \\

Let us list the main findings of this paper:

1. There is an instant of time before the general cosmological singularity in Bianchi IX model,
after which a fulfilment of the condition \eqref{strong} guarantees that it will hold until the singularity is reached.

2. If this condition held before the said instant of time, it will be broken.

3. The smaller the value $C$, which characterizes the non-diagonality of the metric, the closer this instant of time is
to the singularity. Note that the special case of the diagonal metric ($C=0$) cannot be obtained from this solution as a
limit case.

4. The asymptotical dynamics is thus applicable only within the time interval \eqref{e108} between the said instant of time and the
singularity.

5. There is no guarantee that a generic BKL solution will necessarily evolve into asymptotical dynamics, although its
stochastic nature makes it a likely scenario.

It is sufficint to validate the correctness of the asymptotical dynamics in classic GRT. It is worth recalling that we are 
based on Einstein's equations and do not take into account the influence of DE, the scalar field, if any, and quantum effects 
near the singularity. In addition let us note that   
the time interval $t_0$ could be of the same order or less than Planck unit of time if the metrics is
``almost diagonal'' and $C$ is small enough. In this case this approximation is non-physical.
One must take this possibility into account when trying to quantize the Bianchi IX model.

\acknowledgments I am grateful to W. Piechocki for drawing my attention to this problem and for his active and valuable discussion of the 
results obtained.

\end{document}